\documentclass[runningheads]{llncs}
\usepackage{caption}
\usepackage{subcaption}
\usepackage{tikz}
\usepackage{changepage}
\usepackage{adjustbox}
\usepackage{mathtools}
\usepackage{float}

\usepackage{graphicx}
\usepackage{color}
\usepackage{amsmath,amssymb}

\newcommand{\pixCM}[0]{{pix2pix$_{C\rightarrow M}$}}
\newcommand{\pixMC}[0]{{pix2pix$_{M\rightarrow C}$}}

\newcommand{\taskMC}[0]{{MR$\rightarrow$CT}}
\newcommand{\taskCM}[0]{{CT$\rightarrow$MR}}

\newcommand*{\mathcolor}{}
\def\mathcolor#1#{\mathcoloraux{#1}}
\newcommand*{\mathcoloraux}[3]{%
  \protect\leavevmode
  \begingroup
    \color#1{#2}#3%
  \endgroup
}

\begin{document}

\title{Bridging the gap between paired and unpaired medical image translation}
%
%
\author{Pauliina Paavilainen\inst{1,2} \and 
Saad Ullah Akram\inst{1,2}\orcidID{0000-0002-2570-5870} \and
Juho Kannala\inst{1}\orcidID{0000-0001-5088-4041}
}
\authorrunning{P. Paavilainen et al.}
%
\institute{Aalto University, Finland\and
MVision AI, Finland\\
\email{\{pauliina.paavilainen, saad.akram\}@mvision.ai}}

\maketitle              
\begin{abstract}
Medical image translation has the potential to reduce the imaging workload, by removing the need to capture some sequences, and to reduce the annotation burden for developing machine learning methods.
GANs have been used successfully to translate images from one domain to another, such as MR to CT. 
At present, paired data (registered MR and CT images) or extra supervision (e.g. segmentation masks) is needed to learn good translation models.
Registering multiple modalities or annotating structures within each of them is a tedious and laborious task. 
Thus, there is a need to develop improved translation methods for unpaired data. 
Here, we introduce modified pix2pix models for tasks \taskCM{} and \taskMC{}, trained with unpaired CT and MR data, and MRCAT pairs generated from the MR scans.
The proposed modifications utilize the paired MR and MRCAT images to ensure good alignment between input and translated images, and unpaired CT images ensure the \taskMC{} model produces realistic-looking CT and \taskCM{} model works well with real CT as input. 
The proposed pix2pix variants outperform baseline pix2pix, pix2pixHD and CycleGAN in terms of FID and KID, and generate more realistic looking CT and MR translations.

\end{abstract}

\section{Introduction}

\begin{figure}[tb!]
\centering
\includegraphics[width=1\textwidth]{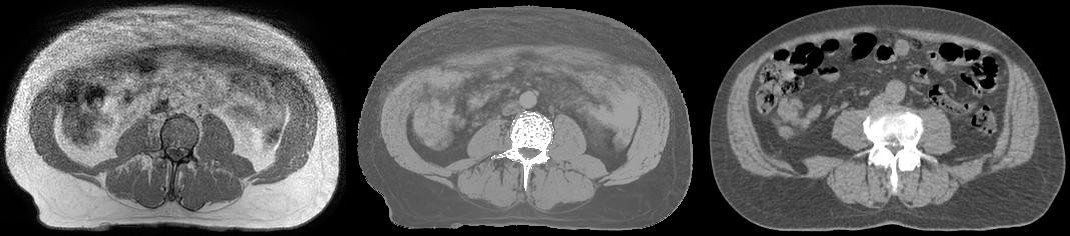}
\begin{minipage}[t]{.33\linewidth}
\centering
\subcaption{MR (DIP)}\label{DIPexample}
\end{minipage}%
\begin{minipage}[t]{.33\linewidth}
\centering
\subcaption{MRCAT}\label{MRCATexample}
\end{minipage}
\begin{minipage}[t]{.33\linewidth}
\centering
\subcaption{CT}\label{CTexample}
\end{minipage}
\caption{MR and MRCAT are voxel-aligned, while CT is from a different patient. 
}
\label{fig:exampleslicesFromDataset}
\end{figure}

Each medical imaging modality captures specific characteristics of the patient.
In many medical applications, complimentary information from multiple modalities can be combined for better diagnosis or treatment.
However, due to limited time, cost and patient safety, not all desired modalities are captured for every patient. 
Medical image translation can play a vital role in many of these scenarios as it can be used to generate non-critical (i.e. the ones which are not needed for fine pattern matching) missing modalities.
One such clinical application is in MR-based radiotherapy, where MR images are used for delineating targets (e.g. tumors) and organs-at-risk (OARs).
However, the clinicians still need CT scans for calculating the dose delivered to OARs and targets.
Since this CT scan is used to compute dose delivered to various structures, the alignment of the patient anatomy is more important than the precise voxel intensities.
There exist few commercial solutions which can be used to generate synthetic CT scans, 
such as Philips' MRCAT. 
However, these translated images do not look very realistic due to quantization artifacts, failure to reduce blurring caused by patient breathing and lack of air cavities, etc. (see Fig.~\ref{MRCATexample}).

Medical image translation can also play a critical role in development of machine learning based image analysis methods.
In many applications, same tasks are performed in different modalities, e.g. in radiotherapy, OAR contouring is done in either CT or MR.
Medical image translation can significantly speed-up the development of these automated methods by reducing the annotation requirements for new modalities/sequences.
A large dataset can be annotated in one modality and then image translation can be used to generate scans of new modalities and annotations copied to synthetic scans.

During recent years, there has been increased interest in using Generative Adversarial Networks (GANs) \cite{goodfellow2014generative} 
in medical image generation. 
The research in the field includes 
low-dose 
CT denoising \cite{wolterink2017generative}, 
MR to CT translation \cite{nie2017medical,wolterink2017deep,emami2018generating,peng2020magnetic}, 
CT to MR translation \cite{jin2019deep}, 
applications in deformable image registration \cite{tanner2018generative} and segmentation \cite{jiang2020psigan}, 
data augmentation \cite{Sandfort2019}, 
PET to CT translation, MR motion correction \cite{armanious2019unsupervised}, and PET denoising \cite{armanious2020medgan} (for review, see \cite{yi2019generative}).

pix2pix \cite{isola2017image} and CycleGAN \cite{zhu2017unpaired} are two popular general-purpose image-to-image translation methods. 
pix2pix requires paired and registered data, while CycleGAN can be trained with unpaired images.
pix2pix and its variants can produce high quality realistic-looking translations, however, capturing voxel-aligned scans or registering scans is a time-consuming task.
In the medical imaging field, there is often lack of paired data, making CycleGAN more suitable method.
However, without any additional constraints, it is difficult to optimize and the translated images may not always have good alignment with input scans.
Many variants have been proposed which impose additional constraints, e.g. mask alignment between both input and output scans \cite{Zhang2018}. 

Besides CycleGAN, another relatively popular unpaired image-to-image translation method is UNsupervised Image-to-image Translation (UNIT) \cite{liu2017unsupervised}, which is a Variational Autoencoder (VAE)-GAN-based method \cite{kingma2013auto,goodfellow2014generative,larsen2016autoencoding}. UNIT has been utilized for T1 to T2 MR and T2 to T1 MR translation \cite{welander2018generative}, PET to CT translation, and MR motion correction \cite{armanious2019unsupervised}. 
Other potential generative models for unpaired image translation include 
Multimodal UNIT (MUNIT) \cite{huang2018multimodal}, Disentangled Representation for Image-to-Image Translation++ (DRIT++) \cite{lee2020drit++}, Multi-Season GAN (MSGAN) \cite{zhang2020msgan}, and StarGAN v2 \cite{choi2020stargan}.

We propose modifications to original pix2pix model for tasks \taskCM{} and \taskMC{} for situations where there is only unpaired CT and MR data.
MR scans can be used to generate voxel-aligned (paired) Magnetic Resonance for Calculating ATtenuation (MRCAT) scans, which look somewhat like CT but are not very realistic and thus not suitable for many tasks (see Fig.~\ref{fig:exampleslicesFromDataset}). 
Our proposed models utilize the alignment information between MR and MRCAT as an auxiliary supervision and produce more realistic CT and MR translations.

\section{Methods}

We propose two pix2pix variants, \pixMC{} and \pixCM{}, for tasks \linebreak
\taskMC{} and \taskCM{}, respectively. 
These models are trained with unpaired CT and MR, and paired MR and MRCAT, which are used as auxiliary supervision to preserve anatomic alignment between input and translated images.
We use U-Net-based \cite{ronneberger2015u} generators and PatchGAN discriminators as in \cite{isola2017image}.

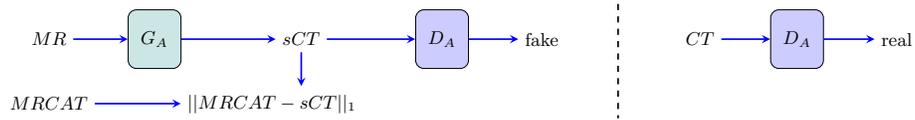
\begin{figure}[!t]
%
\centering
\begin{adjustwidth*}{-1.3em}{0em}%
\scalebox{0.78}{
%
\tikzset{arrow/.style={-stealth, thick, draw=black!80!black}}%
\usetikzlibrary{matrix}%
\begin{tikzpicture}[ampersand replacement=\&]
    \node[] (MRin) at (-0.5,0) {$MR$};
    \node[rectangle, rounded corners, draw, fill=teal!20, minimum height=1cm, minimum width=0.9cm] at (1.3,0) (GA) {$G_A$};
    \node[] at (3.8,0) (fakeCTfromr) {$sCT$};
    \node[] at (-0.5,-1.1) (MRCAT) {$MRCAT$};
    \node[] at (3.3,-1.1) (mrcatL1) {$||MRCAT - sCT||_1$};
    \node[rectangle, rounded corners, draw, fill=blue!20, minimum height=1cm, minimum width=0.9cm] at (6.2,0) (DA1) {$D_{A}$};
    \node[] at (7.9,0) (Fake1) {fake};
    \node[rectangle, rounded corners, draw, fill=blue!20, minimum height=1cm, minimum width=0.9cm] at (12.25,0) (DA12) {$D_{A}$};
    \node[] at (10.6,0) (realCT) {$CT$};
    \node[] at (13.95,0) (Real1) {real};
    \draw[thick,dashed] (9.2,0.6) -- (9.2,-1.35);
    \path[arrow,draw=blue!100!blue,fill=blue!100!blue] 
    (MRin) edge (GA)
    (fakeCTfromr) edge (DA1)
    (GA) edge (fakeCTfromr)
    (realCT) edge (DA12)
    (DA1) edge (Fake1)
    (DA12) edge (Real1);
    \path[arrow,draw=blue!100!blue,fill=blue!100!blue] 
    (MRCAT) edge (mrcatL1)
    (fakeCTfromr) edge ([xshift=0.5cm,yshift=0.3cm]mrcatL1)
    ;%
\end{tikzpicture}%
}%
\end{adjustwidth*}%
\caption{
\pixMC{} has a generator $G_A$, which generates sCT from MR, and a discriminator $D_A$, which distinguishes between real CT and sCT.
$L_{1}$ loss between MRCAT (pair of MR) and sCT is used as an auxiliary supervision.
}%
\label{fig:Pix2pixA_model_schematic}
\end{figure}

\subsubsection{\pixMC{} (\taskMC{}):}
\pixMC{} consists of one generator $G_A$ and an unconditional discriminator $D_A$ (Fig. \ref{fig:Pix2pixA_model_schematic}). 
The generator is trained to generate synthetic CT (sCT=$G_A(MR)$) from real MR input, while the discriminator $D_A$ is trained to classify between the sCT and real CT.  \pixMC{} has an unconditional GAN objective, $\mathcal{L}_{GAN}(G_A, D_A)$, as the conditional GAN (cGAN) objective cannot be used due to lack of paired CT and MR. 
Following \cite{goodfellow2014generative,isola2017image}, 
$D_{A}$ is trained to maximize this objective, while $G_A$ is trained to maximize $\log (D_A(G_A(MR)))$.
    
\begin{equation}
\mathcal{L}_{GAN}(G_A, D_A) =
\mathbb{E}_{CT} [\log D_A(CT)] +
\mathbb{E}_{MR}[\log (1 - D_A(G_A(MR)))]
\label{eq:pixmcgan}
\end{equation}

In addition, $G_A$ is trained to minimize an $L_1$ loss between MRCAT and the generated sCT, $\mathcal{L}_{L_1}(G_A)$. 
Note that we do not have ground truth CTs for the MRs, which is why we use MRCATs in the $L_1$ loss. 
The $L_1$ loss plays similar role as cGAN objective in original pix2pix \cite{isola2017image} as it encourages the sCTs to be aligned with the input MRs.
Our target distribution is the distribution of real CTs instead of MRCATs, which is why the MRCATs are only used as an auxiliary supervision. 

\begin{equation}
\mathcal{L}_{L_1}(G_A) = \mathbb{E}_{MRCAT,MR} [||MRCAT - G_A(MR) ||_1]
\end{equation}

The full training objective of \pixMC{} is

\begin{equation}
\mathcal{L}(G_A, D_A) = \mathcal{L}_{GAN}(G_A, D_A) + \lambda \mathcal{L}_{L_1}(G_A)
\end{equation}

\subsubsection{\pixCM{} (\taskCM{}):}

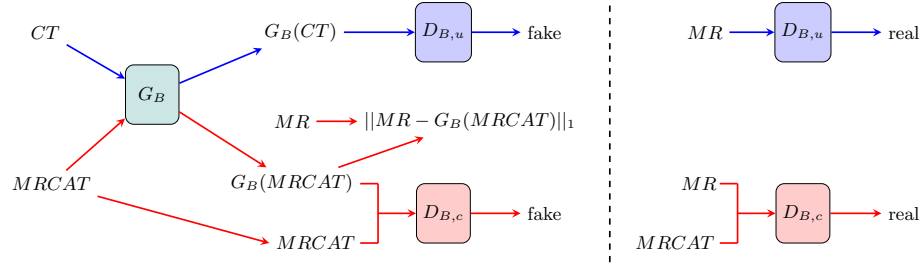
\begin{figure}[!t]
%
\centering%
\begin{adjustwidth*}{-0.4em}{-0.4em}%
\scalebox{0.79}{
\tikzset{arrow/.style={-stealth, thick, draw=black!80!black}}%
\usetikzlibrary{matrix}%
\begin{tikzpicture}[ampersand replacement=\&]
    \node[] (CTin) at (-0.5,0.05) {$CT$};
    \node[rectangle, rounded corners, draw, fill=teal!20, minimum height=1cm, minimum width=0.9cm] at (1.3,-1.0) (GB) {$G_B$};
    \node[] at (3.8,0.05) (fakeMRfromct) {$G_B(CT)$};
    \node[rectangle, rounded corners, draw, fill=blue!20, minimum height=1cm] at (6.2,0.05) (DB1) {$D_{B,u}$};
    \node[] at (7.9,0.05) (Fake1) {fake};
    \node[rectangle, rounded corners, draw, fill=blue!20, minimum height=1cm] at (12.25,0.05) (DB12) {$D_{B,u}$};
    \node[] at (10.6,0.05) (realMR) {$MR$};
    \node[] at (13.95,0.05) (Real1) {real};
    \draw[thick,dashed] (9.0,0.5) -- (9.0,-3.85);
    \node[] (MRCATin) at (-0.4,-2.5) {$MRCAT$};
    \node[] at (3.65,-2.5) (fakeMRfrommrcat) {$G_B(MRCAT)$};
    \node[] (MRCATin2) at (4.05,-3.5) {$MRCAT$};
    \node[] (MRin) at (3.65,-1.45) {$MR$};
    \node[] (MRL1loss) at (6.65,-1.45) {$||MR - G_B(MRCAT)||_1$};
    \node[rectangle, rounded corners, draw, fill=red!20, minimum height=1cm] at (6.2,-3.0) (DB2) {$D_{B,c}$};
    \node[] at (7.9,-3.0) (Fake2) {fake};
    \draw[thick,draw=red!100!red,fill=red!100!red] (4.8,-2.5) -- (5.1,-2.5);
    \draw[thick,draw=red!100!red,fill=red!100!red] (5.1,-3.5) -- (4.8,-3.5);
    \draw[thick,draw=red!100!red,fill=red!100!red] (5.1,-2.5) -- (5.1,-3.5);
    \draw[arrow, draw=red!100!red,fill=red!100!red] (5.1,-3.0) -- (DB2.west);
    \path[arrow,draw=red!100!red,fill=red!100!red] (MRCATin) -- (MRCATin2.west);
    \node[rectangle, rounded corners, draw, fill=red!20, minimum height=1cm] at (12.25,-3.0) (DB22) {$D_{B,c}$};
    \node[] at (10.5,-2.5) (realMR2) {$MR$};
    \node[] at (10.1,-3.5) (realMRCAT) {$MRCAT$};
    \node[] at (13.95,-3.0) (Real2) {real};
    \draw[thick,draw=red!100!red,fill=red!100!red] (10.85,-2.5) -- (11.15,-2.5);
    \draw[thick,draw=red!100!red,fill=red!100!red] (10.85,-3.5) -- (11.15,-3.5);
    \draw[thick,draw=red!100!red,fill=red!100!red] (11.15,-2.5) -- (11.15,-3.5);
    \path[arrow,draw=red!100!red,fill=red!100!red] (11.15,-3.0) -- (DB22.west);
    \path[arrow,draw=blue!100!blue,fill=blue!100!blue] 
    (CTin) edge (GB)
    (fakeMRfromct) edge (DB1)
    (GB) edge (fakeMRfromct)
    (realMR) edge (DB12)
    (DB1) edge (Fake1)
    (DB12) edge (Real1)
    ;
    \path[arrow,draw=red!100!red,fill=red!100!red] 
    (MRCATin) edge (GB)
    (GB) edge (fakeMRfrommrcat)
    (DB2) edge (Fake2)
    (DB22) edge (Real2);
    \path[arrow,draw=red!100!red,fill=red!100!red] 
    (MRin) edge (MRL1loss)
    (fakeMRfrommrcat) edge (MRL1loss)
    ;%
\end{tikzpicture}%
}%
\end{adjustwidth*}%
\caption{\pixCM{} has a generator $G_B$, for producing sMR, and two discriminators, $D_{B,u}$ and $D_{B,c}$. 
$D_{B,c}$, is conditioned on the input MRCAT, and learns to classify between $G_B(MRCAT)$ and real MR. $D_{B,u}$ distinguishes between $G_B(CT)$ and real MR.
$\mathcolor{blue}{\boldsymbol{\rightarrow}}$: path for CT input. $\mathcolor{red}{\boldsymbol{\rightarrow}}$: path for MRCAT input.
}%
\label{fig:Pix2pixB_model}
\end{figure}

\pixCM{} consists of one generator $G_B$ and two discriminators $D_{B,u}$ and $D_{B,c}$. 
The generator takes either real CT, with probability of 0.5, or MRCAT as input during training, and it is trained to generate synthetic MR (sMR) images from both input modalities. 
The unconditional discriminator $D_{B,u}$ is trained to classify between the sMR generated from real CT, $G_B(CT)$, and real MR. 
The conditional discriminator $D_{B,c}$ is conditioned on MRCAT, and is trained to classify between the synthetic MR generated from MRCAT, $G_B(MRCAT)$, and real MR.

\pixCM{} has two GAN objectives: $\mathcal{L}_{GAN}(G_B, D_{B,u})$ and $\mathcal{L}_{cGAN}(G_B, D_{B,c})$. 
Following \cite{goodfellow2014generative,isola2017image}, $D_{B,u}$ and $D_{B,c}$ are trained to maximize  $\mathcal{L}_{GAN}(G_B, D_{B,u})$ and $\mathcal{L}_{cGAN}(G_B, D_{B,c})$, respectively, while $G_B$ is trained to maximize
$\log (D_{B,u}(G_B(CT)))$ and 
$\log (D_{B,c}(MRCAT, G_B(MRCAT)))$.

\begin{align}
    \mathcal{L}_{GAN}(G_B, D_{B,u}) = &
    \mathbb{E}_{MR} [\log D_{B,u}(MR)] +
    \mathbb{E}_{CT}[\log (1 - D_{B,u}(G_B(CT)))]
\end{align}
\begin{align}
    \mathcal{L}_{cGAN}(G_B, D_{B,c}) = &
    \mathbb{E}_{MRCAT,MR} [\log D_{B,c}(MRCAT,MR)] \nonumber \\
    & +
    \mathbb{E}_{MRCAT}[\log (1 - D_{B,c}(MRCAT, G_B(MRCAT)))]
\end{align}

    In addition, $G$ is trained to minimize the $L_1$ loss between MR and the sMR generated from MRCAT, $\mathcal{L}_{L_1}(G_B)$.
    Since we do not have paired CT and MR, we cannot compute $L_1$ loss between $G_B(CT)$ and MR. While our primary goal is \taskCM{} translation, the introduction of MRCAT inputs allows us to use the alignment information between the MRCAT and MR pairs to encourage the generator to produce sMR that is aligned with the input. 
    
    \begin{equation}
    \mathcal{L}_{L_1}(G_B) = \mathbb{E}_{MR,MRCAT} [||MR - G_B(MRCAT) ||_1]
    \end{equation}
    
    The full training objective of \pixCM{} is as follows

\begin{equation}
\mathcal{L}(G_B, D_{B,u}, D_{B,c}) = \mathcal{L}_{GAN}(G_B, D_{B,u}) + \mathcal{L}_{cGAN}(G_B, D_{B,c}) + \lambda \mathcal{L}_{L_1}(G_B)
\end{equation}

\subsubsection{Training details:}
We use random horizontal flip, random zoom (scale: $0.6-1.4$) and random crop as data augmentations. 
In ablation experiments, we use down-sampled (by a factor of 4) images.
We utilize the code \cite{pix2pix,pix2pixhd} provided by the authors \cite{isola2017image,zhu2017unpaired,wang2018high}. 
We use instance normalization, and a batch size of 8 for pix2pixHD, and batch size of 16 for the other models. 
All models are trained for 50 epochs, and 
13K
iterations per epoch. 
We use Adam optimizer with constant learning rate (LR) ($0.0002$) for the first 30 epochs and with linearly decaying LR from $0.0002$ to zero over the last 20 epochs.

For \pixMC{}{} and \pixCM{}, we use the vanilla GAN loss (the cross-entropy objective) like in original pix2pix \cite{isola2017image} (Table \ref{tabtrainData}). 
For \pixCM{} we use $\lambda=100$, and discriminator receptive field (RF) size $70\times70$ in ablation (low resolution) experiments and $286\times286$ in our final model. For \pixMC{}, we use $\lambda=50$, and discriminator RF size $70\times70$.
Since CycleGAN with the default $70\times70$ discriminators fails, we use a stronger baseline CycleGAN with discriminator RF size $142\times142$.

\section{Experiments}

\subsubsection{Dataset:}
The dataset contains 51 pairs of Dixon-In-Phase (DIP) MR and \mbox{MRCAT} scans, and 220 unpaired CT scans of prostate cancer patients, treated with radiotherapy, from Turku University Hospital. 
The scans were randomly split into training and evaluation set.
The number of training MR/MRCAT/CT scans is 41/41/179, and the number of evaluation MR/CT scans is 10/41.

\subsubsection{Evaluation Metrics:}
We use Kernel Inception Distance (KID) \cite{binkowski2018demystifying} and Fréchet Inception Distance (FID) \cite{heusel2017gans} between real and translated images as the evaluation metrics. 
They measure the distance between the distribution of the generated images and the target distribution of real images, and lower values indicate better performance. 
In addition, we use DICE coefficient between automatically segmented \cite{diagnostics10110959} structures (e.g. Body, Femurs, Prostate, etc.) of input and translated images to evaluate the anatomic alignment and quality of translations.
For task \taskMC{}, we also compute the mean absolute HU intensity difference (HU-Dif) between mean HU value for each segmented ROI in real and translated CTs.

\subsection{Comparison with baselines}

We compare our models, \pixMC{} and \pixCM{}, with pix2pix \cite{isola2017image} and pix2pixHD \cite{wang2018high}, and CycleGAN \cite{zhu2017unpaired}. 
Table~\ref{tabtrainData} provides the details of the models and training data.

\begin{table}[!tb]
\centering
\caption{Overview of the models. 
Paired data refers to MR and MRCAT pairs. 
CTs are unpaired with MRs/MRCATs. 
}\label{tabtrainData}
\scalebox{0.93}{
\begin{tabular}{l|l|l|l|l|l|l} 
Model & Generator & GAN mode & Training data &  \#CT & \#MR & \#MRCAT \\
\hline
CycleGAN & unet\_512 & lsgan & unpaired & 179 & 41 & 0 \\
pix2pixHD & global & lsgan & paired & 0 & 41 & 41 \\
pix2pix & unet\_512 & vanilla & paired & 0 & 41 & 41 \\
\pixMC{}/\pixCM{} & unet\_512 & vanilla & unpaired \& paired & 179 & 41 & 41 \\
\hline
\end{tabular}
}
\end{table}

\subsubsection{\taskMC{}:}

\begin{figure}[!b]
\includegraphics[
width=\linewidth
, keepaspectratio=true,trim={0 0 0 4cm},clip]{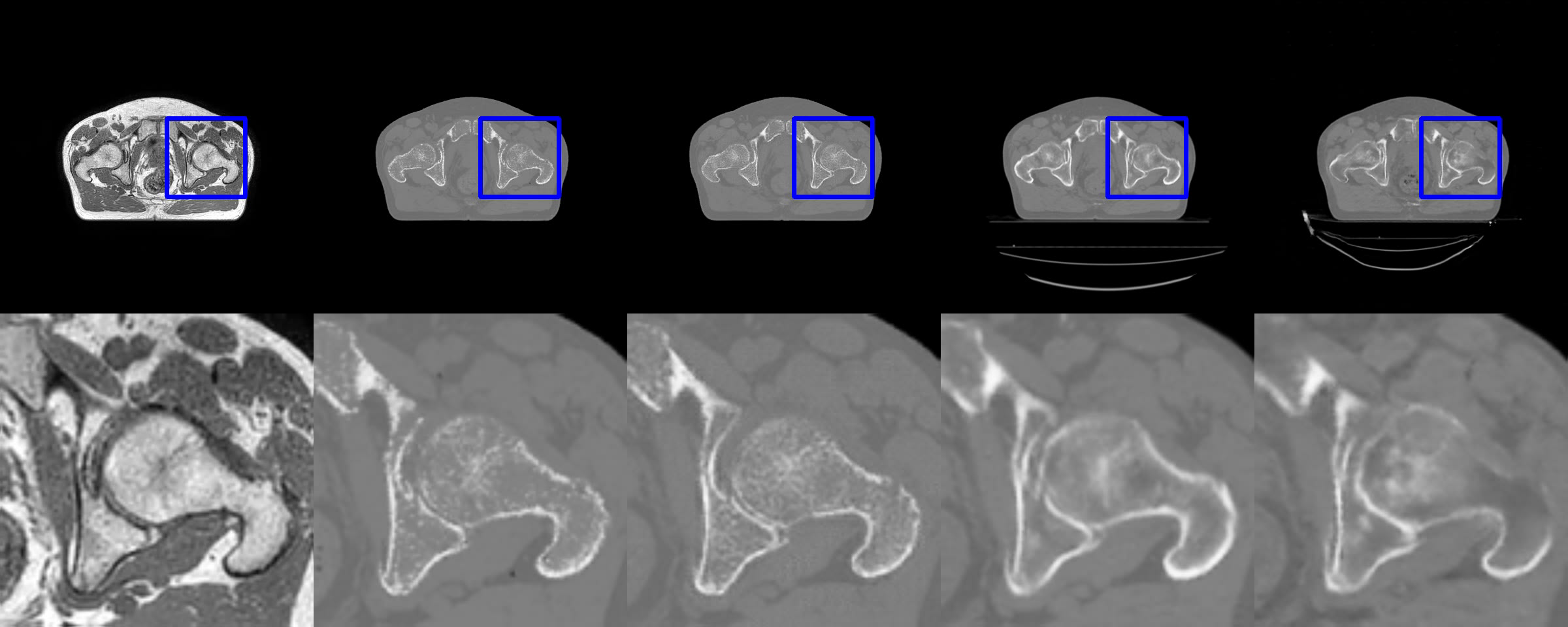}
\begin{minipage}[t]{.18\linewidth}
\centering
\subcaption{Input}\label{apixa}
\end{minipage}%
\begin{minipage}[t]{.2\linewidth}
\centering
\subcaption{pix2pix}\label{bpixa}
\end{minipage}
\begin{minipage}[t]{.2\linewidth}
\centering
\subcaption{pix2pixHD}\label{cpixa}
\end{minipage}
\begin{minipage}[t]{.2\linewidth}
\centering
\subcaption{\pixMC{}}\label{dpixa}
\end{minipage}
\begin{minipage}[t]{.2\linewidth}
\centering
\subcaption{CycleGAN}\label{epixa}
\end{minipage}%
\caption{sCTs produced by \pixMC{} and the baselines. First row: Slices from complete scans. Second row: cropped slices.}
\label{fig:highres_e1_Pix2pixA_and_baselines}
\end{figure}

Table~\ref{tab:taskMC} shows that \pixMC{} outperformed all other methods in terms of FID and KID, indicating that its translated images better resembled real CT.
It had slightly worse DICE (standard deviation is much higher than the difference) compared to pix2pix and pix2pixHD, which can partly be explained by the fewer artifacts produced by these models.
However, since pix2pix and pix2pixHD were trained using MRCAT as their target, their predictions had similar limitations as the MRCAT, e.g. clear quantization artifacts were present (see bones in Fig. \ref{fig:highres_e1_Pix2pixA_and_baselines}). 
pix2pix and pix2pixHD had high FID and KID, primarily due to absence of the patient couch in the translated CT.
CycleGAN translations had small misalignment with inputs and some moderate artifacts, lowering its DICE score. 
\pixMC{} had some prominent artifacts, primarily in the bottom few slices, these might have been caused by the slight difference in the field-of-view of MR and CT datasets. 
\pixMC{} and CycleGAN generated air cavities and hallucinated patient tables.

\begin{table}[!t]
\centering
\caption{\taskMC{} translation: Performance comparison.
}\label{tab:taskMC}
\begin{tabular}{l|l|l|l|l|l|l|l}
Model & \multicolumn{3}{l|}{Training data} & $FID_{CT,sCT}$ & $KID_{CT,sCT}$ & DICE & HU-Dif\\
 & CT & MR & MRCAT & & & \\
\hline
CycleGAN & \checkmark & \checkmark &   & 42.8 & 0.019 & 0.80$\pm$0.20 & \textbf{15.5}\\
pix2pixHD &   & \checkmark & \checkmark & 121.9 & 0.118 & \textbf{0.91$\pm$0.10} & 23.2\\ 
pix2pix &   & \checkmark & \checkmark & 122.6 & 0.119 & \textbf{0.91$\pm$0.12} & 23.0\\
\pixMC{} & \checkmark & \checkmark & \checkmark & {\bf 34.3} & {\bf 0.009} & 0.88$\pm$0.14 & 16.1\\
\hline
\end{tabular}
\end{table}

\subsubsection{\taskCM{}:}

\begin{table}[!b]
\centering
\caption{\taskCM{} translation: Performance comparison.
}\label{tab:taskCM}
\begin{tabular}{l|l|l|l|l|l|l}
Model & \multicolumn{3}{l|}{Training data} & $FID_{MR,sMR}$ & $KID_{MR,sMR}$ & DICE\\
 & CT & MR & MRCAT & & \\
\hline
CycleGAN & \checkmark & \checkmark &    & 55.9 & 0.029 & 0.58$\pm$0.29 \\
pix2pixHD &    & \checkmark & \checkmark & 127.8 & 0.122 & 0.81$\pm$0.17 \\
pix2pix &    & \checkmark & \checkmark & 90.8 & 0.086 & 0.81$\pm$0.16\\
\pixCM{} & \checkmark & \checkmark & \checkmark & {\bf 45.3} & {\bf 0.021} & \textbf{0.83$\pm$0.15}\\
\hline
\end{tabular}
\end{table}

\begin{figure}[!b]
\centering
\includegraphics[
width=\linewidth
, keepaspectratio=true,trim={0 0 0 3cm},clip]{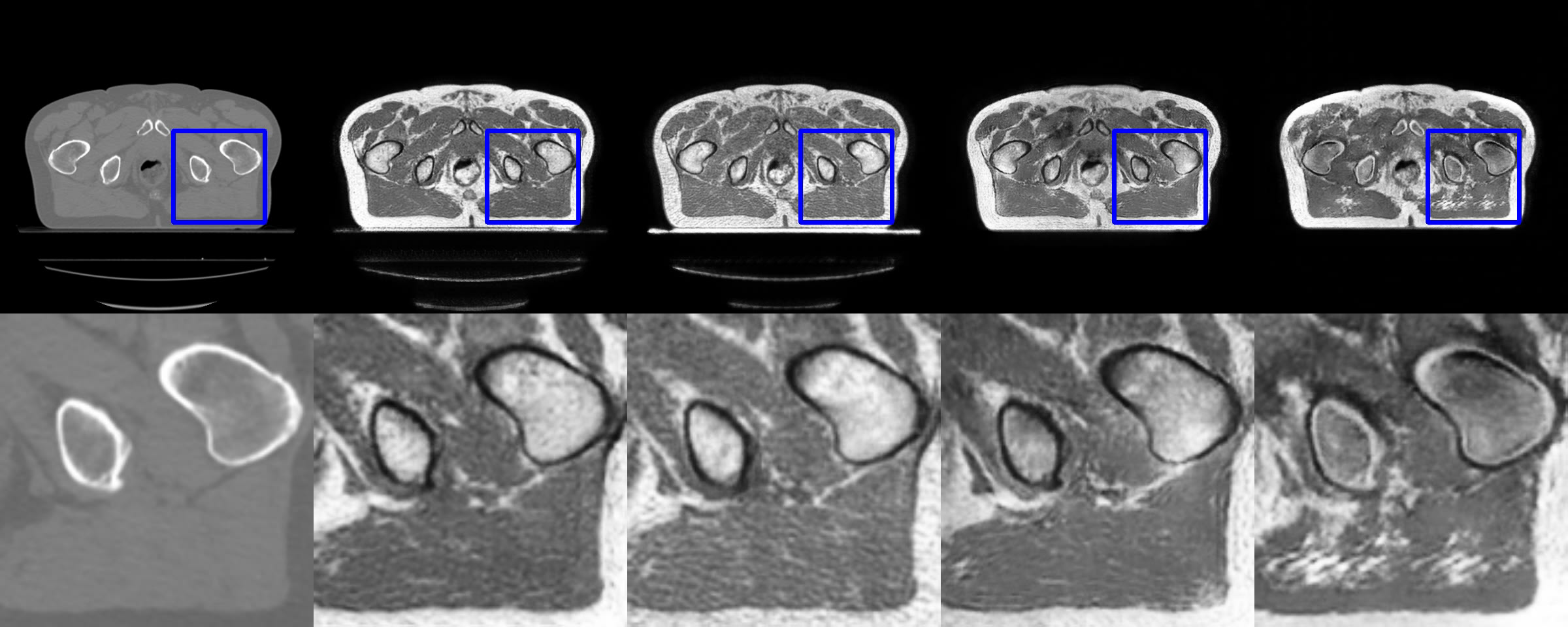}
\begin{minipage}[t]{.18\linewidth}
\centering
\subcaption{Input}\label{e2a1}
\end{minipage}%
\begin{minipage}[t]{.2\linewidth}
\centering
\subcaption{pix2pix}\label{e2b1}
\end{minipage}
\begin{minipage}[t]{.2\linewidth}
\centering
\subcaption{pix2pixHD}\label{e2c1}
\end{minipage}
\begin{minipage}[t]{.2\linewidth}
\centering
\subcaption{\pixCM{}}\label{e2d1}
\end{minipage}
\begin{minipage}[t]{.19\linewidth}
\centering
\subcaption{CycleGAN}\label{sMRcycleGAN} 
\end{minipage}%
\caption{sMRs generated by \pixCM{} and the baselines. First row: Slices from complete scans. Second row: cropped slices.
}
\label{fig:highres_e1_Pix2pixB_and_baselines}
\end{figure}

\pixCM{} had the best performance in terms of FID, KID and DICE, as shown in Table~\ref{tab:taskCM}.
pix2pix and pix2pixHD produced relatively good translations of the patient anatomy but due to their failure to ignore the couch (see Fig.~\ref{fig:highres_e1_Pix2pixB_and_baselines}), visible in CT, their FID and KID values were high. 
CycleGAN had the worst translations, with large artifacts in some parts of the body (see Fig.~\ref{sMRcycleGAN}), 
frequently causing large segmentation failures and leading to very low DICE. 
\pixCM{} produced less artifacts and more realistic sMRs, however, some sMR slices had a small misalignment with inputs near the couch.

\subsection{Ablation studies}\label{AblationSection}

\subsubsection{\pixMC{} objective:} 
When the training objective included both $L_1$ and GAN loss, the translations looked realistic and the performance of \pixMC{} was better in terms of FID, KID and DICE scores compared to the experiment with only $L_1$ objective or only GAN objective (Table \ref{tab1}). 
When only GAN objective was used, the generated sCTs had poor alignment with the input MRs.

\begin{table}[!tb]
\centering
\caption{Training objectives for \pixMC{} 
in low resolution. $\lambda = 50.$}\label{tab1}
\begin{tabular}{l|l|l|l|l|l|l}
Objective & \multicolumn{3}{l|}{Training data} & $FID_{CT,sCT}$ & $KID_{CT,sCT}$ & DICE\\
 & CT & MR & MRCAT & & & \\
\hline
$L_1$ & & \checkmark & \checkmark & 147.4 & 0.146 & 0.72\\
GAN & \checkmark & \checkmark & & 67.3 & 0.032 & 0.23 \\
GAN + $\lambda $$L_1$ & \checkmark & \checkmark & \checkmark & {\bf 39.4} & {\bf 0.014} & \textbf{0.73}\\
\hline
\end{tabular}
\end{table}

\subsubsection{\pixCM{} objective:}

\begin{table}[!tb]
\centering
\caption{Training objectives for \pixCM{} %
in low resolution. $\lambda=100$. 
}\label{tab2}
\begin{tabular}{l|l|l|l|l|l|l} 
Objective & \multicolumn{3}{l|}{Training data} & $FID_{MR,sMR}$ & $KID_{MR,sMR}$ & DICE \\
 & CT & MR & MRCAT & & & \\
\hline
GAN & \checkmark & \checkmark & & 40.0 & 0.015 & 0.24 \\
GAN+cGAN & \checkmark & \checkmark & \checkmark & 32.8 & 0.013 & 0.52 \\ %
GAN+cGAN+$\lambda L_1$ & \checkmark & \checkmark & \checkmark & {\bf 30.0} & \textbf{0.012} & \textbf{0.57}\\ %
\hline
\end{tabular}
\end{table}

When only CT inputs were used with a GAN objective (see $\mathcolor{blue}{\boldsymbol{\rightarrow}}$ path in Fig.~\ref{fig:Pix2pixB_model}), the generated sMR images were not well aligned with the inputs. 
The translations were better aligned when the cGAN objective was included, i.e.  with a probability of 0.5 either CT or MRCAT input was used. 
Inclusion of the $L_{1}$ objective (between sMR generated from MRCAT and real MR) with GAN+cGAN produced the best results in terms of FID, KID and DICE (see Table~\ref{tab2}).

\section{Conclusion}

Our results show that \taskCM{} and \taskMC{} translation with unpaired CT and MR, using the MR and MRCAT pairs as an auxiliary supervision, produces more realistic translated CT and MR images. 
This additional supervision reduces artifacts and improves alignment between the input and the translated images. 
The proposed \pixMC{} and \pixCM{}, outperformed the baseline pix2pix, pix2pixHD and CycleGAN, in terms of FID and KID scores.

\pixMC{} and \pixCM{}, like other GAN-based methods, may be useful in producing realistic-looking translated images for research purposes.
Since these methods can hallucinate features in images \cite{cohen2018distribution}, they require extensive validation of their image quality and fidelity before clinical use. 
It remains as task for future research to develop improved translation methods and design quantitative metrics which better capture the quality of translations.

\clearpage
\bibliographystyle{splncs04}
\bibliography{mybibliography}

\end{document}